\documentclass[letterpaper,conference]{IEEEtran}
\IEEEoverridecommandlockouts

\usepackage{amsmath,amssymb,amsfonts}
\usepackage{bm}
\usepackage{mathtools}

\def\BibTeX{{\rm B\kern-.05em{\sc i\kern-.025em b}\kern-.08em
    T\kern-.1667em\lower.7ex\hbox{E}\kern-.125emX}}

\usepackage{caption}
\usepackage{xcolor}
\usepackage{graphicx}
\usepackage{xspace}
\usepackage{wrapfig}
\usepackage{subcaption}
\usepackage{tikz}
\usepackage{hyperref}
\hypersetup{
  colorlinks=true,
  linkcolor=red,
  filecolor=magenta,      
  urlcolor=cyan,
  citecolor=blue,
}
\usepackage{authblk}
\usepackage{csvsimple}

\usepackage[linesnumbered,ruled,vlined]{algorithm2e}
\DontPrintSemicolon

\SetKwComment{Comment}{\color{green!50!black}// }{}

\SetKwProg{Function}{function}{}{}

\makeatletter
\DeclareRobustCommand\onedot{\futurelet\@let@token\@onedot}
\def\@onedot{\ifx\@let@token.\else.\null\fi\xspace}

\def\eg{\textit{e.g}\onedot} 
\def\ie{\textit{i.e}\onedot}

\foreach \x in {a,...,z}{
	\expandafter\xdef\csname bf\x \endcsname{\noexpand\ensuremath{\noexpand\mathbf{\x}}}
	\expandafter\xdef\csname bm\x \endcsname{\noexpand\ensuremath{\noexpand\boldsymbol{\x}}}
}
\foreach \x in {A,...,Z}{
	\expandafter\xdef\csname bs\x \endcsname{\noexpand\ensuremath{\noexpand\boldsymbol{\x}}}
	\expandafter\xdef\csname bf\x \endcsname{\noexpand\ensuremath{\noexpand\mathbf{\x}}}
	\expandafter\xdef\csname bb\x \endcsname{\noexpand\ensuremath{\noexpand\mathbb{\x}}}
	\expandafter\xdef\csname ds\x \endcsname{\noexpand\ensuremath{\noexpand\mathds{\x}}}
	\expandafter\xdef\csname cal\x \endcsname{\noexpand\ensuremath{\noexpand\mathcal{\x}}}
}

\def\herm{\mathsf{H}}

\renewcommand{\phi}{\varphi}

\def\real{\mathrm{Re}}

\renewcommand{\epsilon}{\varepsilon}

\DeclareGraphicsExtensions{.eps,.pdf,.png,.jpg}

\DeclareMathOperator*{\argmin}{arg\,min}

\definecolor{myred}{RGB}{153,0,0}			
\definecolor{mygreen}{RGB}{51,153,0}	    
\definecolor{mygold}{RGB}{140,110,10}		
\definecolor{mylgray}{RGB}{80,80,80}		
\definecolor{mygray}{RGB}{51,51,51}		    
\definecolor{myash}{RGB}{100,100,100}		
\definecolor{mypurple}{RGB}{80,20,120}		
\definecolor{myblue}{RGB}{18,75,126}

\newcommand{\red}[1]{\textcolor{myred}{#1}}

\makeatletter
\newcommand*\l@chapterinfo{\@nodottedtocline{0}{0.0em}{1.5em}}
\newcommand*\l@sectioninfo{\@nodottedtocline{1}{1.5em}{2.3em}}
\newcommand*\l@subsectioninfo{\@nodottedtocline{2}{3.8em}{3.2em}}
\newcommand*\l@subsubsectioninfo{\@nodottedtocline{3}{7.0em}{4.1em}}
\newcommand*\l@paragraphinfo{\@nodottedtocline{4}{10em}{5em}}
\newcommand*\l@subparagraphinfo{\@nodottedtocline{5}{12em}{6em}}

\def\@nodottedtocline#1#2#3#4#5{%
	\ifnum #1>\c@tocdepth \else
	\vskip \z@ \@plus.2\p@
	{\leftskip #2\relax \rightskip \@tocrmarg \parfillskip -\rightskip
		\parindent #2\relax\@afterindenttrue
		\interlinepenalty\@M
		\leavevmode
		\@tempdima #3\relax
		\advance\leftskip \@tempdima \null\nobreak\hskip -\leftskip
		{#4}\nobreak
		\leaders\hbox{$\m@th
			\mkern \@dotsep mu\hbox{\,}\mkern \@dotsep
			mu$}\hfill
		\nobreak
		\hb@xt@\@pnumwidth{\hfil\normalfont \normalcolor }%
		\par}%
	\fi}

\makeatother

\def\chapterinfo#1{%
	\addcontentsline{toc}{chapterinfo}{%
		\noexpand\numberline{}#1}%
}
\def\sectioninfo#1{%
	\addcontentsline{toc}{sectioninfo}{%
		\noexpand\numberline{}#1}%
}
\def\subsectioninfo#1{%
	\addcontentsline{toc}{subsectioninfo}{%
		\noexpand\numberline{}#1}%
}
        
\begin{document}
	
	\title{Time and covariance smoothing \\ for restoration of bivariate signals
		{\thanks{The authors acknowledge support from the ANR
		project “Chaire IA Sherlock” ANR-20-CHIA-0031-01, the programme d'investissements d'avenir ANR-16-IDEX0004 ULNE and Region Hauts-de-France. This research has been supported in whole or in part, by the Agence	Nationale de la Recherche (ANR) under project ANR-21-CE48-0013-03. In the interest of open access publication, the author/rights holder applies a CC-BY open access license to any article/manuscript accepted for publication (AAM) resulting from this submission. The corresponding author: {firstname.lastname}@univ-lille.fr: firstname:$\{$yusuf-yigit$\}$, lastname: $\{$pilavci$\}$.}}}
	
	\author[1]{Yusuf Yi\u{g}it P\.{I}LAVCI}
	\author[2]{Pierre PALUD}
	\author[3]{Julien FLAMANT}
	\author[1]{\\Pierre-Antoine THOUVENIN}
	\author[1]{J\'er\'emie BOULANGER}
	\author[1]{Pierre CHAINAIS}

	\affil[1]{Univ. Lille, CNRS, Centrale Lille, UMR 9189 CRIStAL, F-59000, Lille, France} 
	\affil[2]{Universite Paris Cit\'e, CNRS, Astroparticule et Cosmologie,	F-75013 Paris, France}
	\affil[3]{Universit\'e de Lorraine, CNRS, CRAN, F-54000 Nancy, France}
	%
	\maketitle
	
	\begin{abstract}
		In many applications and physical phenomena, bivariate signals are polarized, \textit{i.e.} they trace an elliptical trajectory over time when viewed in the 2D planes of their two components. 
		The smooth evolution of this elliptical trajectory, called polarization ellipse, is highly informative to solve ill-posed inverse problems involving bivariate signals where the signal is collected through indirect, noisy or incomplete measurements. 
		This work proposes a novel formulation and an efficient algorithm for reconstructing bivariate signals with polarization regularization. 
		The proposed formulation leverages the compact representation of polarization through the instantaneous covariance matrices. 
		To address the resulting quartic optimization problem, we propose a well-suited parameter splitting strategy which leads to an efficient iterative algorithm (alternating direction method of multipliers (ADMM)) with convex subproblems at each iteration.
		The performance of the proposed method is illustrated on numerical synthetic data experiments.  
	\end{abstract}
	
	\section{Introduction}
	\sectioninfo{\red{
			- Bivariate signal processing, applications in cosmology, seismology, acoustics, and neuroscience. \checkmark \\			
			- Inverse problems for bivariate signals, with a focus on their polarization properties. \checkmark \\
			- Mention on Stokes but more covariance matrix \checkmark \\
			- Polarization regularization via Jones polarizer (Cyril), Eusipco paper. I'll look more if there are more stuff related \checkmark }
	}

	A myriad of examples involving electromagnetic waves, gravitational waves, seismic waves, underwater acoustics signals or EEG signals falls into the realm of bivariate signal processing (BSP), \ie, the study of signals with two correlated components~\cite{flamant2018phdthesis}. 
	In multivariate signal processing, BSP is a special case where many useful representations are accessible.
	Several seminal works~\cite{flamant2019time,flamant2018complete,miron2023quaternions} have studied the definition and practical use of quaternions, covariance matrices and Stokes parameters (widely used in optics) to interpret the \textit{polarization} properties, 
	\ie the properties related to the elliptical trajectory described by the bivariate signal in the 2D plane.
	Solving inverse problems involving bivariate signals, such as signal restoration, relies on these representations to formulate polarization priors into regularization~\cite{pilavci2024denoising,cano2022mathematical}.
    They often lead to non-linear, typically \textcolor{black}{quartic}, formulations.
	As a result, the associated regularizations are often non-convex functions. 
	Previously, the Jones polarizer matrix has been used to impose a predefined polarization for reconstructing gravitational waves~\cite{cano2023contrainte}.
	This method is very effective when the polarization is known,~\eg, circular or linear. 
	However, knowing the polarization drastically limits the range of application and excludes cases of time varying polarization.
	Closer to this work,~\cite{pilavci2024denoising} proposes two formulations to impose smoothness both in the time and Stokes domains. 
	The numerical results show that the best reconstruction results are obtained by regularizing in both domains. 
	However, the solutions to these non-convex formulations are given by gradient descent based algorithms, which are not necessarily well adapted for the problem at hand.  
	In this work, we propose a new formulation of the signal reconstruction problem from noisy measurements.
	In doing so, we use the instantaneous covariance matrices~\cite{flamant2018phdthesis}, and propose a parameter splitting strategy handled with ADMM optimization algorithm~\cite{boyd2011distributed} to solve the corresponding quartic optimization problem. 
	
	The rest of the paper is organized as follows.  
	Section~\ref{sec:problem-desc} gives the formulation of the problem with various regularizations.
	Section~\ref{sec:method} presents the proposed method based on an ADMM optimization algorithm. 
	Section~\ref{sec:num-illust} illustrates and discusses the performance and runtime of the method. 
	Concluding remarks are reported in Section~\ref{sec:conc}.
	
	\section{Problem description}
	\label{sec:problem-desc}
	
    \subsection{Forward model and problem formulation}
	\subsectioninfo{\red{
		- Forward model with generic $\Phi$ \\
		- MLE formulation (no explicit solution)  \\}
	}
	
    Given  a bivariate signal $\bfX = \big[ u[n], v[n] \big]_{1 \leq n \leq N}\in\mathbb{R}^{N\times2}$ composed of $N$ samples, 
	the observation model for the noisy signal $\mathbf{Y} = (\mathbf{y}_d)_{1 \leq d \leq D} \in \mathbb{R}^{N \times D}$ measured in $D$ channels is
	\begin{equation}
		\bfY = \Phi(\bfX) + \boldsymbol{\epsilon} \in \mathbb{R}^{N\times D},
		\label{eq1}
	\end{equation}
	where $\Phi$ is a known \textcolor{black}{linear} operator, and the noise $\boldsymbol{\epsilon}=[\boldsymbol{\epsilon}_{d}]_{1 \leq d \leq D}$ \textcolor{black}{has} stationary Gaussian components $\boldsymbol{\epsilon}_d \sim \mathcal{N}(\mathbf{0},\boldsymbol{\Xi}_d)$ \textcolor{black}{with known covariance matrices $\boldsymbol{\Xi}_d\in\mathbb{R}^{N\times N}$.} For simplicity, one can consider $\boldsymbol{\Xi}_d = \mathbf{I}_N$.
	The generic linear transformation $\Phi$ appears in various applications where the bivariate signal is not directly observed \textcolor{black}{and} reconstructed from measurements \textcolor{black}{over} $D$ channels, such as gravitational signal processing~\cite{cano2022mathematical}. 
 	To solve this reconstruction problem, one needs to impose certain priors on the reconstructed signal $\bfX$ to obtain better reconstruction performance. These priors can be generically formulated as 
	\begin{equation}
		\bfX^{\star} \in \argmin_{\bfX \in \bbR^{N\times 2}}f(\bfX,{\bfY}) + \lambda_1g_1(\bfX) + \lambda_2g_2(\bfX),
		\label{eq:opt-prob}
	\end{equation}
	with data-fidelity term $f(\bfX,\bfY)=\Vert\Phi(\bfX) -\bfY\Vert_F^2$. 
	The regularization functions $g_1$ and $g_2$ respectively reflect the prior on the 2 components of the bivariate signal and the geometric properties, \textcolor{black}{with corresponding hyperparameters} $\lambda_1$ and $\lambda_2$.

	\subsection{Towards relevant regularizations}
	A typical prior on individual components of $\bfX$ is that each component has smooth variations over time~\cite{tikhonov1977solutions,mallat1999wavelet}.
	This is often modeled by penalizing the amplitude of the time gradient of $\bfX$, which can be formulated $g_1(\bfX) = \lambda_1\Vert\bfD\bfX\Vert_F^2$ with $\bfD$ the finite difference operator. The squared Frobenius norm reflects the assumption of Gaussian increments between two successive samples.
	This regularization leads to smooth evolution of each component, but does not ensure that both the signal and the polarization evolve smoothly over time.  
	In bivariate signals, the polarization ellipse typically changes much slower than the frequency content of the signal~\cite{cano2022mathematical}. Thus, the reconstructed signal is assumed to have a smooth polarization evolution. 
	In this work, we propose to use the instantaneous covariance matrices as a representation of the polarization ellipse.

	Starting from the analytic signal $\bfX_{a}=\bfH\bfX = \big[u_a[n],v_a[n] \big]_{n\in\{1,\dots,N\}}\in\mathbb{C}^{N\times 2}$ where $\bfH\in\bbC^{N\times N}$ filters out the negative frequencies~\cite{marple1999computing}, the instantaneous covariance matrix can be defined as   
	\begin{equation}
		\begin{split}
			\boldsymbol{\Sigma}[n] &= {\bfX_{a}[n]}^\herm\bfX_{a}[n]= \bfX^\herm\bfH^\herm\bfe_n\bfe_n^\top\bfH\bfX \\
			&= \begin{bmatrix}
				|u_a[n]|^2 & u_a[n]^\herm v_a[n] \\
				v_a[n]^\herm u_a[n] & |v_a[n]|^2  
			\end{bmatrix}, \\ 
		\end{split}
	\end{equation}
	where $\bfe_n$ is the $n$-th canonical basis vector of \textcolor{black}{$\mathbb{R}^N$}. Here, $u_a[n]$ and $v_a[n]$ are the analytic versions of $u[n]$ and $v[n]$. 
	The eigendecomposition of $\boldsymbol{\Sigma}[n]$ is parametrized by the instantaneous polarization ellipse parameters such as its size, eccentricity and orientation~\cite{flamant2018phdthesis}. 
	As a result, {smoothing $\boldsymbol{\Sigma}[n]$} leads to smooth evolution of the polarization ellipses through time. 
	
	
    \textcolor{black}{To promote a smooth polarization evolution, we propose to penalize the evolution of the covariance matrices using finite differences with}
	\begin{align}
		g_2(\bfX) &= \sum_{n=2}^N\Vert \boldsymbol{\Sigma}[n] - \boldsymbol{\Sigma}[n-1] \Vert_F^2 \nonumber \\ 
				  &= \sum_{n=2}^N\Vert \bfX^\top\bfH^\herm\bfe_n \bfe_n^\top  \bfH\bfX - \bfX^\top\bfH^\herm\bfe_{n-1} \bfe_{n-1}^\top \bfH\bfX  \Vert_F^2 \nonumber \\ 
				  &= \sum_{n=2}^N\Vert \bfX^\top\bfH^\herm\bfJ_n\bfH\bfX  \Vert_F^2,
        \label{eq:polarization_smoothness_regularization}
	\end{align}
	where $\bfJ_n=(\bfe_n \bfe_n^\top - \bfe_{n-1} \bfe_{n-1}^\top)$. 
	{The Euclidean distance does not necessarily take the geometry of covariance matrices into account, yet remains as a good approximation locally~\cite{peyre2009geodesic}.}
		
	The proposed method, called Time And COvariance Smoothing (TACOS), consists in solving the optimization problem in~\eqref{eq:opt-prob} with these proposals for $g_1$ and $g_2$. 
	This problem is differentiable, quartic due to the covariance regularization term $g_2$ so potentially not convex, and does not admit a closed-form solution.
	In the next section, a parameter splitting strategy is used to solve the problem while exploiting its structure. 

	\section{Proposed algorithm for TACOS}
	\label{sec:method}

	\subsection{Parameter splitting and ADMM}
    %
    \textcolor{black}{We propose to apply parameter splitting to the quartic term~\eqref{eq:polarization_smoothness_regularization} so that alternating the optimization between the primal variable and a splitting variable involves easier-to-solve convex sub-problems.}
	To do so, we define another parameter $\bfZ=\bfH\bfX$ and rewrite
	\begin{equation}
		\bfX^\star \in \argmin_{\substack{\bfX \in \bbR^{N\times 2} \\ \bfZ=\bfH\bfX\in\bbC^{N\times2}}}f(\bfX,{\bfY}) + \lambda_1g_1(\bfX) + \lambda_2 \widetilde{g}_2(\bfX,\bfZ),
		\label{eq:splitted-problem}
	\end{equation}
	where $\widetilde{g}_2(\bfX,\bfZ) = \sum_{n=2}^N\Vert \bfX^\top\bfH^\herm\bfJ_n\bfZ \Vert_F^2$.
	The problem~\eqref{eq:splitted-problem} can be addressed with an ADMM algorithm~\cite{boyd2011distributed}.
	The constraint $\bfZ=\bfH\bfX$ is imposed by minimizing
	\begin{equation}
		\begin{split}
		L(\bfX,\bfZ,\bfU) &= f(\bfX,{\bfY}) +  \lambda_1g_1(\bfX) + \lambda_2 \widetilde{g}_2(\bfX,\bfZ) \\  
		& \quad + \frac{\rho}{2} \big(\Vert\bfH\bfX - \bfZ + \bfU\Vert_F^2 - \Vert \bfU\Vert_F^2 \big),
		\end{split}
	\end{equation} 
	where $\rho>0$ adjusts the balance between the constraint and minimization of the objective function. 
	Each iteration $l \in \mathbb{N}$ of ADMM is defined by the update steps
	\begin{subequations}
		\begin{align}
			\bfX^{(l+1)} &= \argmin_{\bfX\in\mathbb{R}^{N\times 2}} L(\bfX,\bfZ^{(l)},\bfU^{(l)}) \\
			\bfZ^{(l+1)} &= \argmin_{\bfZ\in\mathbb{C}^{N\times 2}} L(\bfX^{(l+1)},\bfZ,\bfU^{(l)}) \\
			\bfU^{(l+1)} &= \bfU^{(l)} + \rho(\bfH\bfX^{(l+1)} - \bfZ^{(l+1)}). 
		\end{align}
		\label{eq:admm-steps}
	\end{subequations}
	The typical convergence criteria for such an algorithm are the primal or the dual error being under a certain threshold~\cite{boyd2011distributed}. 
	The parameter $\rho>0$ can have a significant effect on the convergence of this algorithm.
    It can be chosen either empirically or with automated strategies~\cite{boyd2011distributed,eckstein1992douglas}, which goes beyond the scope of this work.  

	\subsection{Detailed update steps}
	\subsectioninfo{The linear system to solve at each step. A sentence added on the most expensive part \checkmark}

	To illustrate each update steps, we consider the following generic linear model
	\begin{equation}
		\forall d\in\{1,\dots,D\},\quad \bfy_d = \bfT_d\bfX\bfr_d + \epsilon_d,
		\label{eq:obs-model}
	\end{equation}
	that is, the observation operator $\Phi$ in~\eqref{eq1} takes the form $\Phi(\bfX) = (\bfT_d \mathbf{X}\bfr_d)_{1 \leq d \leq D}$, with $\bfT_d$ and $\bfr_d$ generic left-hand and right-hand side transformations.
	Many applications fall under this generic linear scheme, notably the reconstruction of gravitational waves from interferometric detectors~\cite{cano2022mathematical}. 
	
	In this case, the first two steps in~\eqref{eq:admm-steps} require solving linear systems of the form of $\bfM_{x}^{(l)}\texttt{vec}(\bfX^{(l+1)})=\bfb_{x}^{(l)}$ and $\bfM_{z}^{(l)}\bfZ^{(l+1)}=\bfb_{z}^{(l)}$, with
	\begin{align}
		\bfM_{x}^{(l)} = &\sum_{d=1}^D(\bfr_d\bfr_d^{\top}\otimes \real\{\bfT_d^\herm\bfT_d\})
		+ [ \bfI_2\otimes (\lambda_1\bfD^\top\bfD \nonumber \\+&\real\{\lambda_2\sum_{n=2}^{N}\bfH^{\herm}\bfJ_n^\top\bfZ^{(l)}(\bfZ^{(l)})^\herm\bfJ_n\bfH\} + 2\rho\bfI)]\nonumber  \\
		\bfb_{x}^{(l)} = &\texttt{vec}\left(\sum_{d=1}^{D} \real\{\bfT_d^\herm\widetilde{\bfy}_d\bfr_d^{\top} +\rho\bfH^{\herm}(\bfZ^{(l)}-\bfU^{(l)})\}\right) \nonumber \\ 
		\bfM_{z}^{(l)} = &\lambda_2\sum_{i=2}^N  \bfJ_n\bfH\bfX^{(l+1)}(\bfX^{(l+1)})^\herm\bfH^\herm \bfJ_n^{\top} +\rho\bfI_N\nonumber \\
		\bfb_{z}^{(l)} = &\rho(\bfH\bfX^{(l+1)}+\bfU^{(l)}),
	\end{align}
	where $\otimes$ is the Kronecker product, the $\texttt{vec}$ operator reshapes an input matrix into a vector by stacking its columns.
	Note that $\bfM_{z}^{(l)}$ is a tridiagonal matrix which can be \textcolor{black}{handled} via Thomas' algorithm~\cite{marsdentexts} in time complexity $\mathcal{O}(N)$. {Moreover, thanks to the identity term, $\bfM_{z}^{(l)}$ is usually well-conditioned.}
	In addition, every operator in $\bfM_{x}^{(l)}$ can be implemented efficiently using well-known algorithms such as the fast Fourier transform~\cite{cooley1965algorithm}. 
	This alleviates the need to store $\bfM_{x}^{(l)}$ as a large matrix for large $N$, while the cost of its product with a vector remains as $\mathcal{O}(N\log N)$. 
	Then, to solve the linear system $\bfM_{x}^{(l)}\bfX=\bfb_{x}^{(l)}$, we adopt the conjugate gradient descent (CGD) algorithm, which accesses $\bfM_{x}^{(l)}$ by only matrix-vector products through $K$ iterations\footnote{The number of iterations $K$ depends on many other variable such as the condition number of $\bfM_{x}^{(l)}$ or the stopping criterion. We leave this study for the future work which is highly connected to the hyperparameter tuning}. \textcolor{black}{CGD is initialized with $\bfX^{(l)}$ to determine $\bfX^{(l+1)}$, and $\bfX^{(0)}$ is initialized randomly}.
    \textcolor{black}{Overall, this part is the most computationally intensive in the algorithm, with complexity $\mathcal{O}(KN\log N)$.}

	\section{Numerical Illustration}
	\label{sec:num-illust}
	\subsection{Experimental setting}
	\subsectioninfo{
		\red{{1/ Explaining simulation model: The signal, the noise, the projection matrix etc. } \checkmark \\ 	
			2/ The numbers of the experiments such as the number of samples, repetitions, varying parameters etc. \checkmark \\
			3/ The performance metrics \checkmark}}	
	To evaluate the performance of the algorithm, synthetic bivariate signals are generated as in~\cite[(1.57)]{flamant2018phdthesis} using random smoothly-varying ellipse parameters. 
	The generated signals are narrow-band, monochromatic and AM-FM-PM signals whose main frequency is contained in the range $[25/N,35/N]$ Hz. 
    
    Synthetic observations are generated in $D = 3$ channels from~\eqref{eq:obs-model}. The entries of $(\bfr_{d})_{1 \leq d \leq D}$ drawn independently from the standard normal distribution.
	To generate the stationary noise $\boldsymbol{\epsilon}_{d}$, first, its amplitude spectral density is generated as $\alpha_d[f]\sim\mathcal{U}(0,\sigma)$ with $\sigma > 0$, and its phase $\phi_d[f]\sim\mathcal{U}(0,2\pi)$ is sampled for all frequencies $f$. The noise is finally generated as $\boldsymbol{\epsilon}_{d}=\mathcal{F}^{-1}\{\boldsymbol{\alpha}_d\exp(\boldsymbol{j}\phi_d)\}$. 
	The operator $\bfT_d=\texttt{diag}(\boldsymbol{\alpha}_d)^{-1}$ can be interpreted as a spectral whitening operator, with $\texttt{diag}(\boldsymbol{\alpha}_d)$ a diagonal matrix whose diagonal is the vector $\boldsymbol{\alpha}_d$. This is often used in the forward model used in gravitational wave restoration~\cite{cano2022mathematical}.
    
	Throughout the simulations, different values have been taken for the scalar $\sigma \in \{10^{-1},\dots,10^{1}\}$ and the number of samples $N\in\{512,1024,4096\}$. Reconstruction results have been averaged over several repetitions. 
	The hyperparameters $\lambda_1\in\{ 10^{-1}, 10^0, 10^1, 10^2, 10^3\}$ and $\lambda_2\in\{ 10^{2}, 10^{3}, 10^4, 10^5, 10^6\}$ are tuned via grid search, while the parameter $\rho$ is set to 1.
	
	The proposed algorithm is compared with the method described in~\cite{pilavci2024denoising}, referred as Stokes and time smoothing (STS).
	This approach relies on smoothing time and Stokes parameters, another representation of polarization. 
	By its definition, STS is not directly adapted to a generic linear forward model and requires noisy measurements of the bivariate signal. 
	To overcome this issue, we feed this algorithm with the MLE solution as input (correspodning to the case $\lambda_1 = \lambda_2=0$, abbreviated $\lambda_{1,2}=0$ in the following).
	
	The maximum number of iterations for both STS is $L=1000$, while TACOS is stopped at $L=100$ or when both primal  $\frac{\Vert\bfZ^{(l)} - \bfH\bfX^{(l)}\Vert_F}{\Vert\bfH\bfX^{(l)}\Vert_F}$ and dual error $\frac{\Vert\bfZ^{(l)} - \bfZ^{(l-1)}\Vert_F}{\Vert\bfZ^{(l-1)}\Vert_F}$ is less than $10^{-3}$. Reconstruction performance is reported in terms of the reconstruction SNR,~\ie r-SNR($\cdot$) = $10\log\frac{\Vert\bfX\Vert_F^2}{\Vert\bfX - \cdot\Vert_F^2}$.
	
    \begin{table}
        \caption{Runtime (in seconds) for varying sample size $N$.}
		\centering
        \begin{filecontents*}{runtime_results.csv}
$N$,MLE,STS,TACOS
512,0.001,5.6,41.3 
1024,0.001,6.2,107.3 
4096,0.01,12.9,1000.0 
\end{filecontents*}
\csvautotabular{runtime_results.csv}
        \label{tab:runtime}
	\end{table}

	\subsection{Results and discussion}
	\subsectioninfo{
		\red{1/ Figs. on illustration and tables on forward bivariate SNR (and maybe runtime)   \\ 	
			2/ A short discussion on how to choose hyperparam and speed-up strategies,\\}}

	\begin{figure}
		\includegraphics[trim={2cm 0 2cm 0},clip,width=1\linewidth]{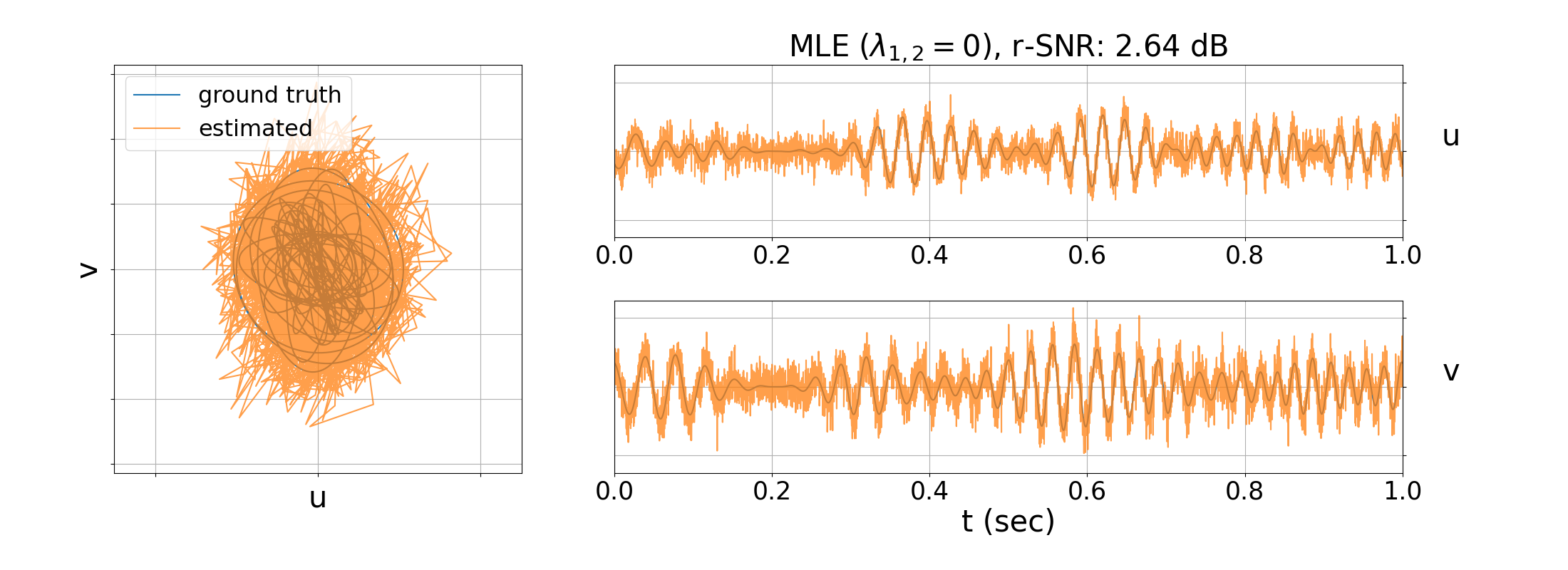}
		\includegraphics[trim={2cm 0 2cm 0},clip,width=1\linewidth]{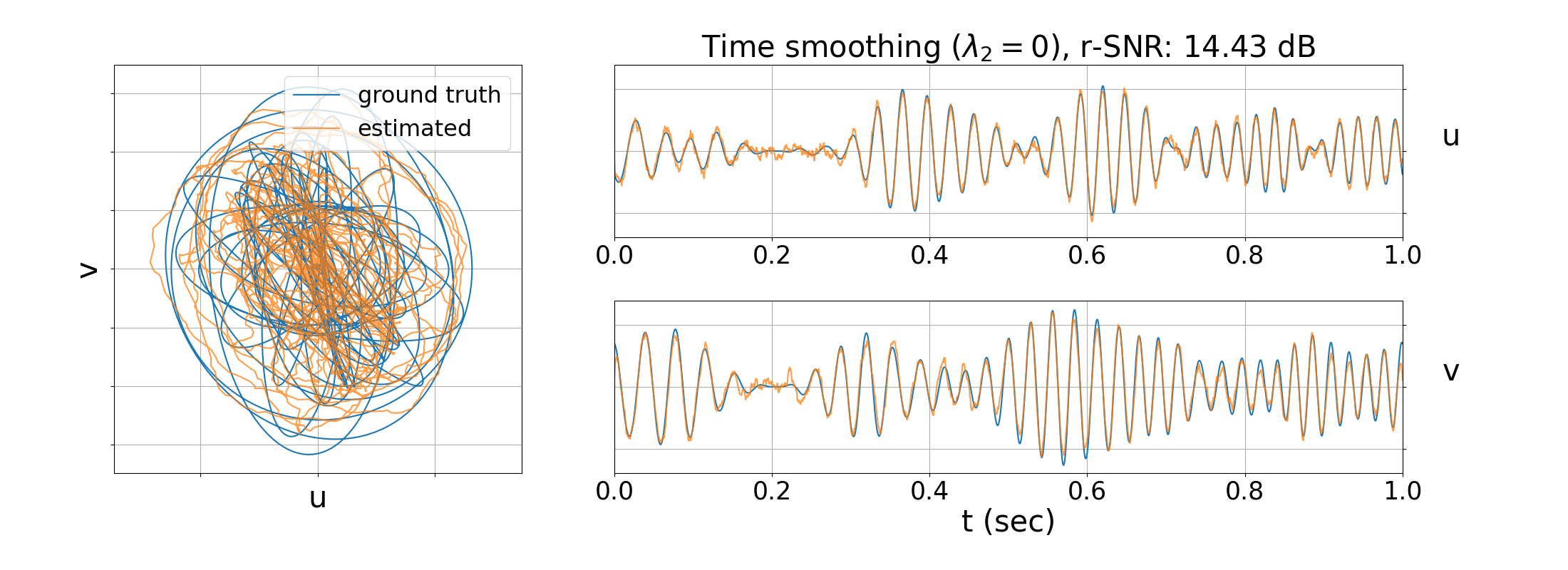}
		\includegraphics[trim={2cm 0 2cm 0},clip,width=1\linewidth]{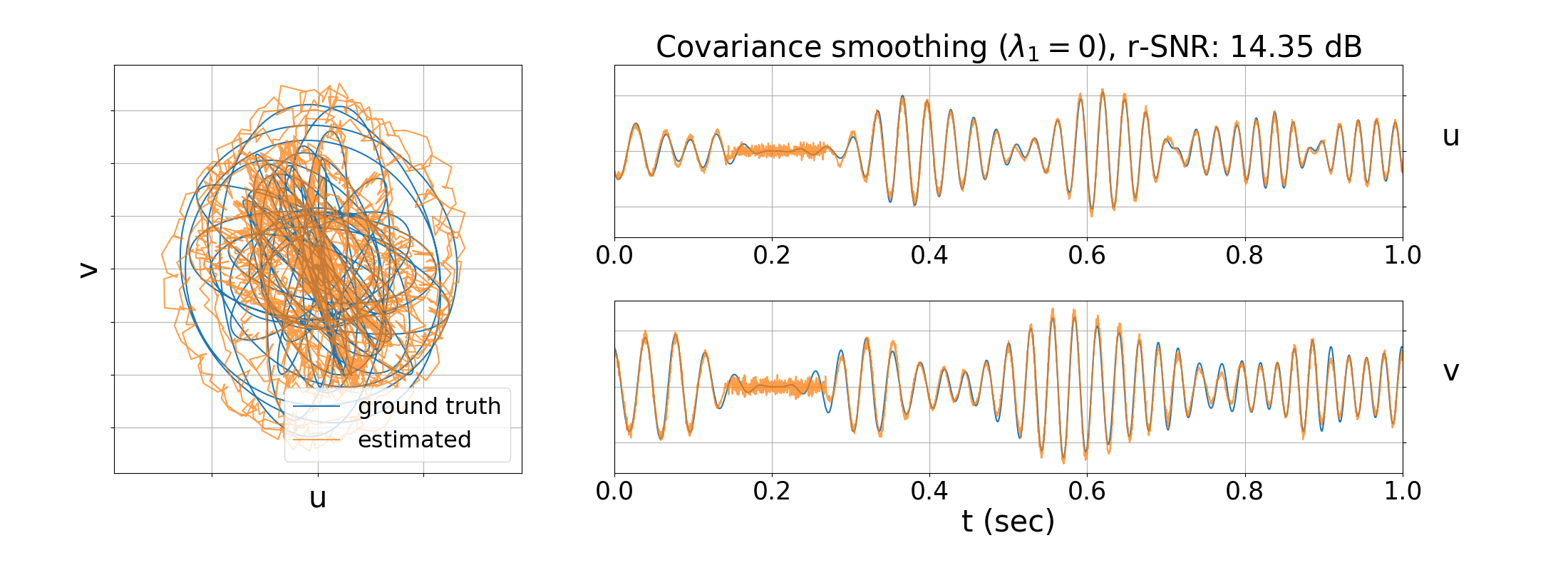}
		\includegraphics[trim={2cm 0 2cm 0},clip,width=1\linewidth]{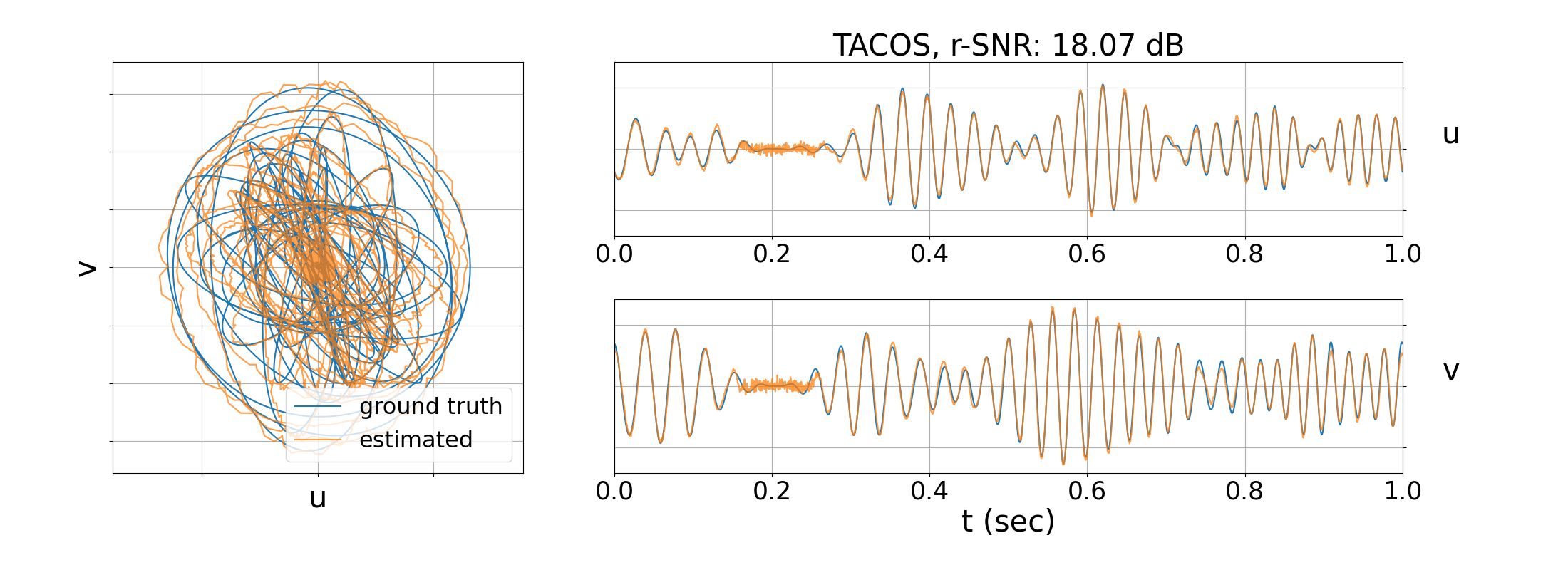}
		\caption{Example reconstruction of a bivariate signal. From top to bottom: MLE~(\ie~$\lambda_1=\lambda_2= 0$) with r-SNR $= 2.64$dB, time smoothing (TS) ($\lambda_{1}>0$, $\lambda_{2}=0$) with r-SNR $=14.43$ dB,
		covariance smoothing (COS) ($\lambda_{1}=0$, $\lambda_{2}>0$) with r-SNR $=14.35$ dB,
		and proposed method ($\lambda_1 >0, \lambda_2>0$) with r-SNR $=18.07$ dB.}
		\label{fig:illustrate-mle-ts}
	\end{figure}
	\begin{figure}
		\centering
		\includegraphics[width=0.92\linewidth]{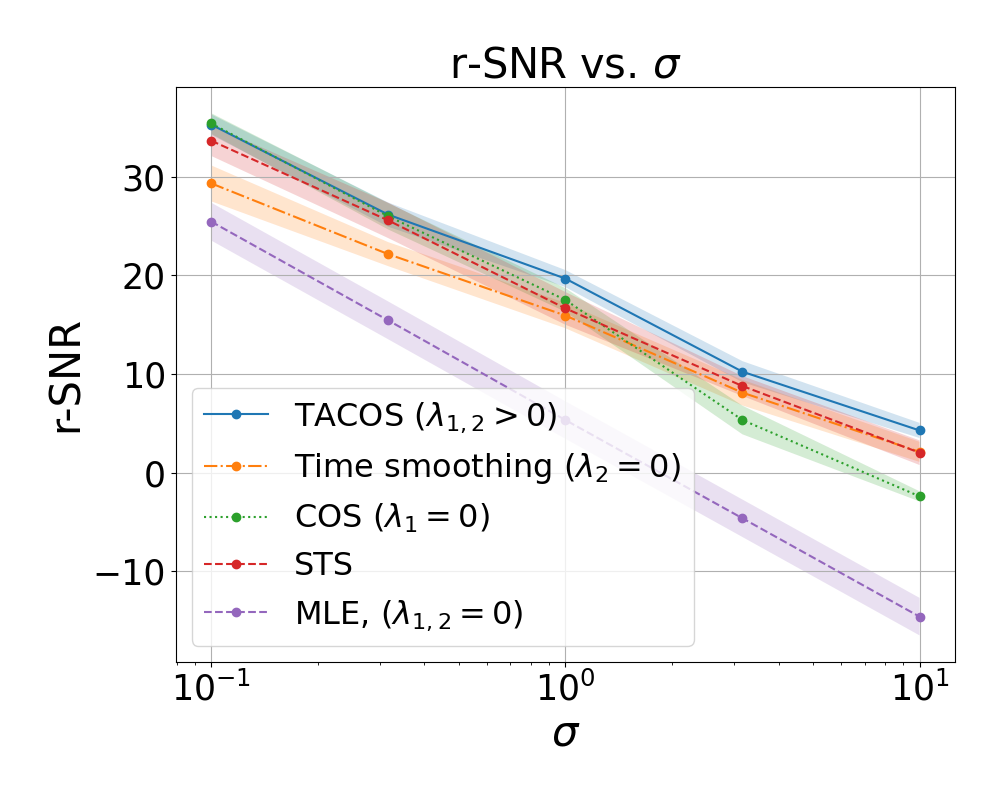}
		\caption{The r-SNR of the bivariate signal vs. the noise scaling factor $\sigma$ for $N=4096$.}
		\label{fig:SNR}
	\end{figure}
	Fig.~\ref{fig:SNR} shows the reconstruction performances of TACOS with different configurations ({$(\lambda_{1}>0,\lambda_{2}=0)$,$(\lambda_{1}=0,\lambda_{2}>0)$ and $(\lambda_{1, 2}>0$)}) and STS. 
	A quick visual comparison confirms that the best results are obtained for TACOS when both time and covariance smoothing are active ($\lambda_{1,2} > 0$). 
	A closer look at Fig.~\ref{fig:illustrate-mle-ts} shows that 
	\textcolor{black}{time-smoothness alone} (r-SNR $=14.43$ dB) does not impose sufficiently smooth elliptical trajectories.
	Similarly, smoothing polarization only (r-SNR $=14.35$ dB) does not yield smooth individual components.
	The best results (r-SNR $=18.07$ dB) are obtained by regularizing the signal in both \textcolor{black}{domains}.  

	The proposed method TACOS performs moderately better than STS as illustrated in Fig.~\ref{fig:SNR} and exploits the structure of the problem with a convenient formulation. 
	On the other hand, the time comparison in Table~\ref{tab:runtime} shows that TACOS takes a longer time compared to STS, especially when $N$ is large. Despite the use of different Python computing backends in the implementation, TACOS still requires a large dense matrix inversion which does not scales properly with the number of samples. The time-consuming step of TACOS is the resolution of the system $\bfM_{x}^{(l)}\texttt{vec}(\bfX)=\bfb_{x}$ at each ADMM iteration $l$, as $\bfM_{x}^{(l)}$ is a block-wise dense matrix.
	
	A different formulation of the optimization problem, \textit{e.g} expressed in terms of the analytic signal, could potentially lead to a different formulation requiring a simpler $\bfM_{x}$ matrix.
	Finding well-suited preconditioner or initialization strategy is another perspective for further acceleration~\cite{saad2003iterative}.

	Regarding the reconstruction performances, it is relatively stable with respect to the hyperparameters: $\lambda_1\in[5, 20]$ and $\lambda_2\in [5\times10^5, 20\times10^5]$ yield minor fluctuations in r-SNR, between $17.5$ and $18.5$ dB.
	Nevertheless, one may look for more efficient ways than a grid search.
    More sophisticated Bayesian optimization algorithms~\cite{pereyra2015maximum,snoek2012practical} can be adapted which is left for the future work on real-life applications.

	\section{Conclusion}
	\sectioninfo{1/ A short conclusion \checkmark, \\
		2/ Possible applications \checkmark} 
	\label{sec:conc}
	In this work, we propose a novel way of formulating polarization priors for bivariate signal reconstruction. 
	This formulation relies on the instantaneous covariance matrices for the polarization/geometric properties. 
	Addressing this formulation within an ADMM algorithm leads to a principled method for denoising polarized bivariate signals.  
	This work presented a numerical validation of the proposed TACOS method on realistic synthetic data.
	Future works will validate the proposed approach on real-life signals, such as the reconstruction of gravitational waves~\cite{cano2022mathematical}.  
	
	
    \bibliographystyle{IEEEtran}
	\bibliography{refs}
\end{document}